\documentclass[epj]{webofc}
\usepackage[utf8]{inputenc}
\usepackage[varg]{txfonts}   
\usepackage{booktabs}
\usepackage{xcolor}
\usepackage{subfigure} 
\definecolor{darkred}{rgb}{0.4,0.0,0.0}
\definecolor{darkgreen}{rgb}{0.0,0.4,0.0}
\definecolor{darkblue}{rgb}{0.0,0.0,0.4}
\usepackage[bookmarks,linktocpage,colorlinks,
    linkcolor = darkred,
    urlcolor  = darkblue,
    citecolor = darkgreen]{hyperref}
\usepackage{subfigure}
\usepackage{xspace} 
\usepackage{mathtools}
\wocname{EPJ Web of Conferences}
\woctitle{Lattice2017}
\newcommand{\rF}{{\ensuremath{F}}\xspace}
\newcommand{\rA}{{\ensuremath{A_2}}\xspace}
\newcommand{\rFs}{\rF{}s\xspace}
\newcommand{\rAs}{\rA{}s\xspace}

\newcommand{\rlF}{{F}}
\newcommand{\rlA}{{A_2}}

\newcommand{\Tr}{\ensuremath{\mathop{\text{Tr}}}}     

\begin{document}
%
\selectlanguage{english}
\title{%
Chiral Transition of SU(4) Gauge Theory with Fermions in Multiple Representations
}
\author{%
\firstname{Venkitesh} \lastname{Ayyar}\inst{1} \and
\firstname{Thomas} \lastname{DeGrand}\inst{1}  \and
\firstname{Daniel~C.}  \lastname{Hackett}\inst{1}%
\thanks{Speaker, \email{daniel.hackett@colorado.edu}} \and
\firstname{William~I.}  \lastname{Jay}\inst{1} \and
\firstname{Ethan~T.}  \lastname{Neil}\inst{1,2}\fnsep \and
\firstname{Yigal}  \lastname{Shamir}\inst{3} \and
\firstname{Benjamin}  \lastname{Svetitsky}\inst{3}
}

\institute{%
Department of Physics, University of Colorado, Boulder, Colorado 80309, USA
\and
RIKEN-BNL Research Center, Brookhaven National Laboratory, Upton, New York 11973, USA
\and
Raymond and Beverly Sackler School of Physics and Astronomy, Tel~Aviv University, 69978 Tel~Aviv, Israel
}
\abstract{%
We report preliminary results on the finite temperature behavior of SU(4) gauge theory with dynamical quarks
 in both the fundamental and two-index antisymmetric representations.
This system is a candidate to present scale separation behavior,
where fermions in different representations condense at different temperature or coupling scales.
Our simulations, however, reveal a single finite-temperature phase transition at which both
 representations deconfine and exhibit chiral restoration.
It appears to be strongly first order.
We compare our results to previous single-representation simulations. We also describe
a Pisarski-Wilczek stability analysis, which suggests that the transition should be first order.
}
\maketitle
\section{Introduction}
\label{sec:intro}
Lattice gauge theories with fermions in multiple representations (``multirep'' theories) provide an arena to test the old 
ideas of tumbling or (for vectorlike systems)  scale separation \cite{Raby:1979my}.
The physical picture is that when a gauge coupling becomes sufficiently strong in the
infrared, a scalar fermion bilinear will form, breaking chiral symmetry.
In a system with multiple representations of fermions, a weaker gauge coupling is needed to drive
condensation for higher representation fermions, since their color charges are greater.
Thus, different representations of fermion may condense at different scales.
Quenched simulations from the early 80's (performed on small lattices with large gauge couplings)
\cite{Kogut:1983sm,Kogut:1982rt,Kogut:1984nq,Kogut:1984sb}
appeared to show such behavior. However, quenching neglects the back-reaction of the fermions
on the gauge dynamics. 
It may also happen that the scales for chiral symmetry breaking and deconfinement are different.
Some old simulations with dynamical fermions (see \cite{Karsch:1998qj})
indicate this behavior. The issue with these systems is that they all
appear to be near or beyond the conformal window (whose precise boundary is still controversial), so they may not be chirally broken at all.
 (See refs.~\cite{DeGrand:2015zxa,ben_review} for reviews.)

Here, we study an SU$(4)$ gauge theory with two flavors of Dirac fermions charged under the fundamental irreducible representation (irrep) \rF (quartet, \textbf{4}) of SU$(4)$; and an additional two flavors of Dirac fermions charged under
 the two-index antisymmetric irrep \rA (sextet, \textbf{6}) of SU$(4)$.
The first- and second-order coefficients of the beta function are negative, signaling that this system is likely to
be an ordinary confining and chirally broken system at zero temperature. Our simulations show that is the case \cite{spectro_paper}.
In these Proceedings, we describe our preliminary results for the finite-temperature phase
 structure of this theory.

\section{Lattice Details}
\label{sec:lattice}
The gauge action used in our simulations is the
 ``nHYP dislocation suppressing'' (NDS) action \cite{DeGrand:2014rwa}.
This is a sum of the usual plaquette term plus an additional term made from smeared links.
We scale all new parameters together, leaving $\beta$ the only free parameter in the gauge sector.
The fermions use the clover action.
The links have nHYP smearing \cite{Hasenfratz:2001hp,Hasenfratz:2007rf}.
The clover coefficient is taken to be $c_{sw}=1$, a choice
known to work well with nHYP smearing in QCD \cite{Bernard:1999kc} and higher-irrep
fermions \cite{Shamir:2010cq}. We have a three-dimensional space of bare parameters: $\beta$
and the hopping parameters $\kappa_4$ for the \rFs and $\kappa_6$ for the \rAs.
Spectroscopic quantities are measured as screening masses, that is, from
correlators projected along one of the spatial directions of the lattice.
When we quote a value for a quark mass, we mean the quantity extracted from the
Axial Ward Identity, or AWI quark mass.

Simulations are performed on $12^3\ times 6$ and $16^3 \times 8$ volumes. The
two $N_t$'s allow us to diagnose whether an observed crossover or transition is a 
finite temperature transition or a bulk one.

To diagnose separately whether each irrep of fermion is confined, we use the fundamental-irrep Polyakov loop and the \rA higher-representation Polyakov loop.
The presence of dynamical fundamental fermions mean neither Polyakov loop is an exact order parameter for the breaking of center symmetry.
Physically, a Polyakov loop in some representation measures the free-energy cost of a static charge of that irrep.
So, we still expect the Polyakov loop to jump from a small value to a large one when the corresponding irrep of charge deconfines.
We thus associate a change in the Polyakov loops with the words ``confinement-deconfinement transition.''

To determine whether chiral symmetry is broken for an irrep, we should check for the presence 
of the corresponding chiral condensate.
Because we are using Wilson fermions the condensates are difficult to access directly, so we
instead use chiral parity doubling as an indirect probe of whether a condensate has formed for either irrep.
In the chirally-restored phase, the GMOR relation ($m_{PS}^2 \propto m_q$ is lost
and the pseudoscalar mass plateaus at $m_P = \pi T$, while 
the parity-partner scalar and pseudoscalar meson states, and
the vector and pseudovector
meson states, become nearly degenerate.
This degeneracy disappears when chiral symmetry is broken.
 
More details on our phase diagnostics are published in a companion proceedings
 \cite{other_thermo_proceedings}.

\section{Results: Antisymmetric-Only Limit}
\label{sec:sextet-only}
To complement our investigation of the full (both fermions dynamical) theory,
 we have also examined the phase structure of the one-representation limiting cases of our theory
 (with either the \rFs or the \rAs infinitely massive and decoupled).
Our group previously studied the \rA-only theory using the same action without the NDS term \cite{DeGrand:2015lna}.
These results allow us to diagnose the effects of the new term.

\begin{figure}
	\centering
	\sidecaption
	\includegraphics[clip,trim=0in 0.1in 0.45in 0.475in,width=3.025in]{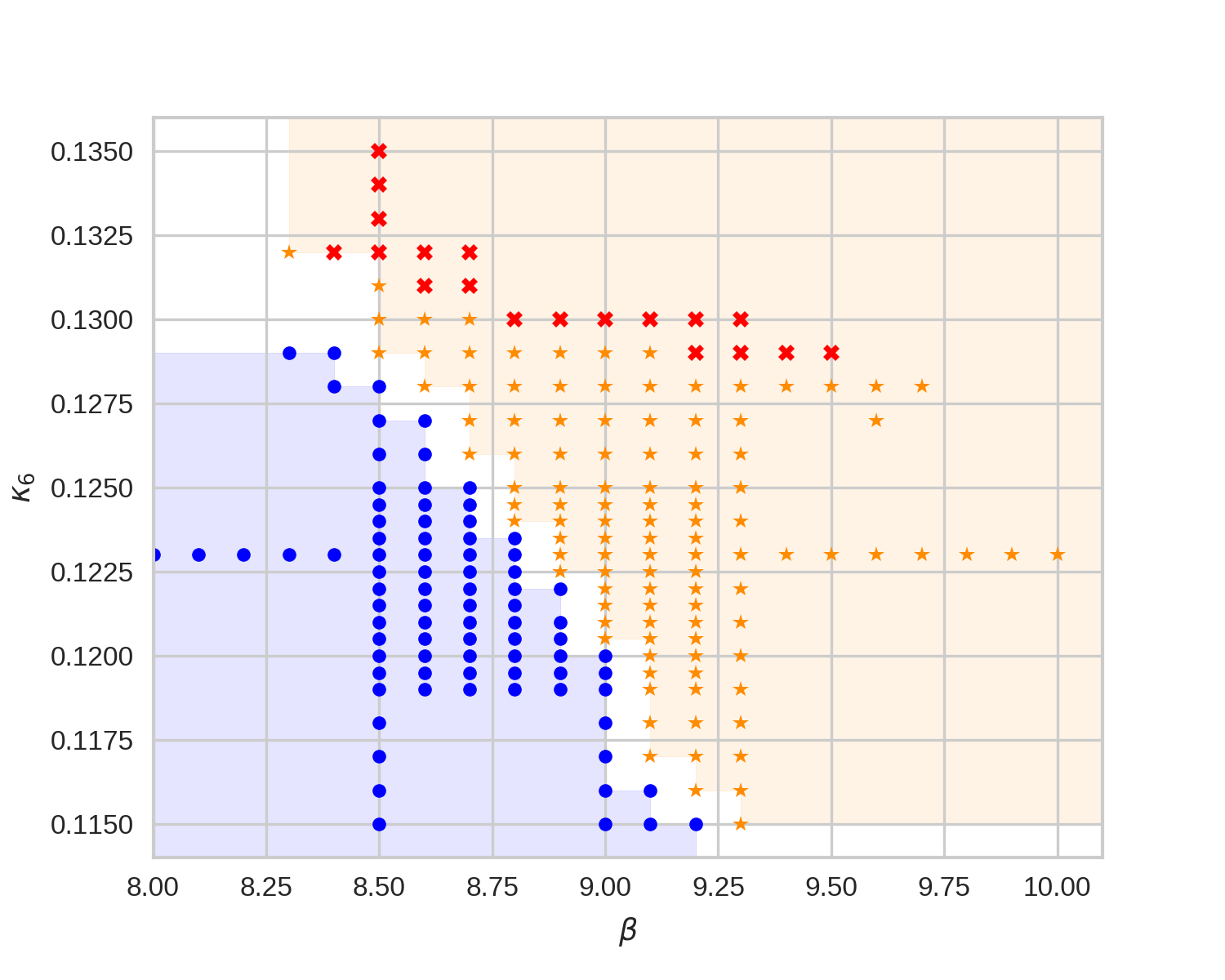}
	\caption{
		Phase diagram for the \rA-only theory with the NDS term, from the present study.
		All data from lattices with $N_t=6$.
		Blue dots indicate confined ensembles where $m_{q6} > 0$. 
		Yellow stars (red Xs) indicate deconfined ensembles with $m_{q6}>0$ ($m_{q6} < 0$).
		Blue regions are unambiguously confined, orange regions are unambiguously deconfined, and white regions are phase-ambiguous (transition lives here).
	}
	\label{fig:a2-phase-diagrams}
\end{figure}

\begin{figure}
	\centering
	\sidecaption
	\includegraphics[width=3in,clip,trim=0.1in 0 0.1in 0]{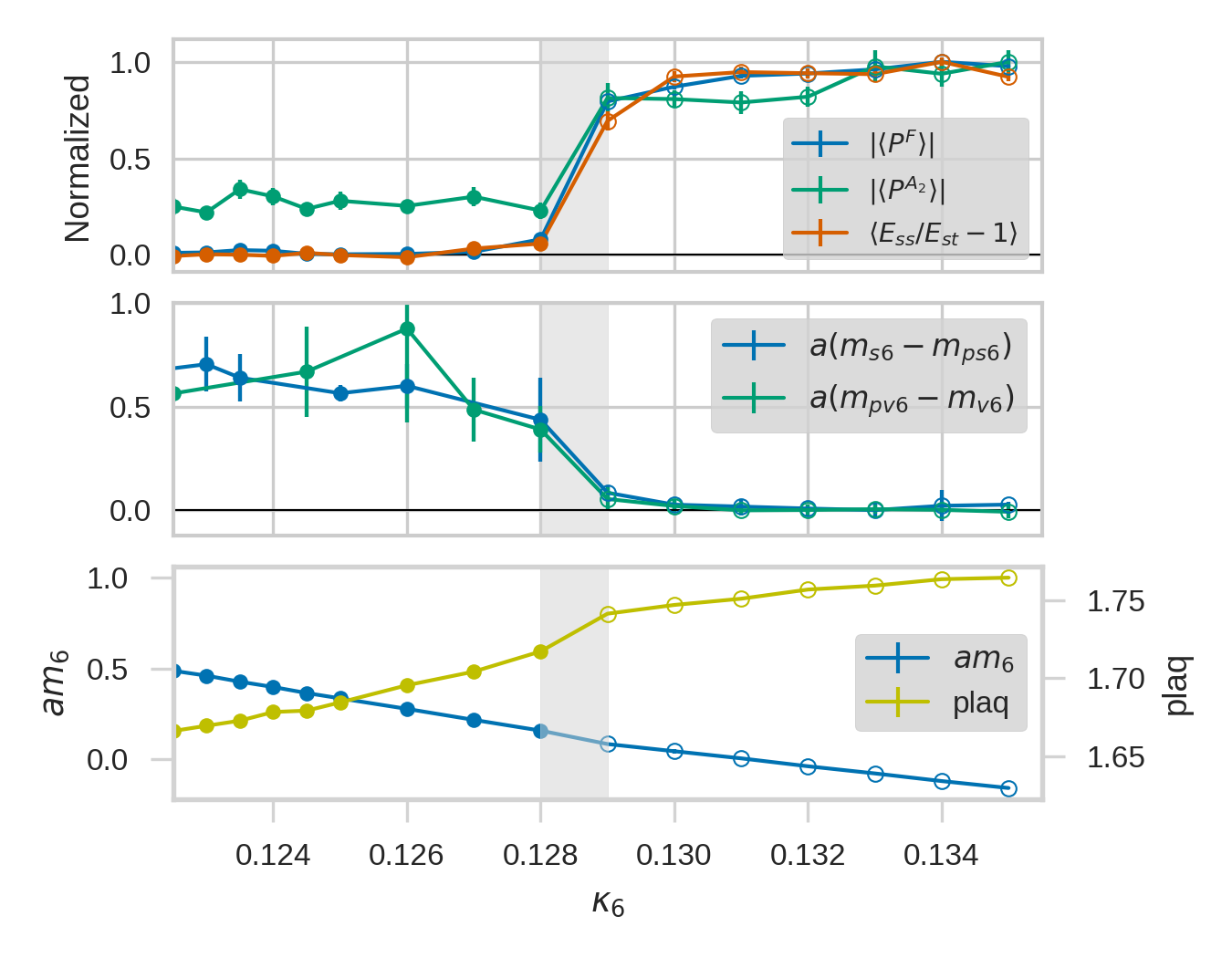}
	\caption{
		Polyakov loops (unflowed) and flow anisotropy (measured at $t/a^2=1$) \cite{other_thermo_proceedings,Datta:2015bzm,Datta:2016kea,Wandelt:2016oym} (top),
		chiral doubling observables (middle),
		and quark mass and plaquette (bottom) 
		varying $\kappa_6$ along a slice of constant $\beta=8.5$ on $N_t=6$ in the \rA-only theory with the NDS action.
	}
	\label{fig:a2_slice}
\end{figure}

Figure~\ref{fig:a2-phase-diagrams} shows our new findings for the phase structure of the \rA-only theory.
We find that the confinement and chiral transitions
coincide everywhere that we have investigated, as seen in typical slices through bare
parameter space as in figure~\ref{fig:a2_slice}. The upper panel shows gauge observables
(Polyakov loops in the two representations and the spatial-temporal anisotropy in the Wilson flow \cite{other_thermo_proceedings,Datta:2015bzm,Datta:2016kea,Wandelt:2016oym}).
The middle panel shows the  scalar-pseudoscalar and vector-axial vector mass differences.
They both show a step at the same value of $\kappa_6$.
The bottom panel shows the AWI quark mass, which is not much affected at the crossover, and the plaquette, which orders.

Our previous study, with a Wilson gauge action, saw an additional first-order bulk transition;
essentially all observables jumped as we crossed it. With the NDS action, we
see no evidence for such a transition where we have looked.

A high order Pisarski-Wilczek-style calculation \cite{Basile:2004wa,Basile:2005hw} finds a 
stable fixed point for this theory and thus predicts that the chiral transition may be second order.
We observe a smooth crossover, consistent with (but not demonstrative of) this prediction.
This is also consistent with the previous study of this system \cite{DeGrand:2015lna}.

\section{Results: Multirep Phase Structure}
\label{sec:multirep}
We collected data for the multirep theory on $12^3 \times 6$ and $16^3\times 8$ lattices
at 21 values of the bare gauge coupling and many (typically $O(10)$) values of $(\kappa_4,\kappa_6)$
per $\beta$ value.
We explored $\beta=7.4$ and $\beta=7.75$ more extensively, with 159 and 236 ensembles respectively.

\begin{figure}
	\centering
	\sidecaption
	\includegraphics[width=3in,clip,trim=0.15in 0 0.1in 0]{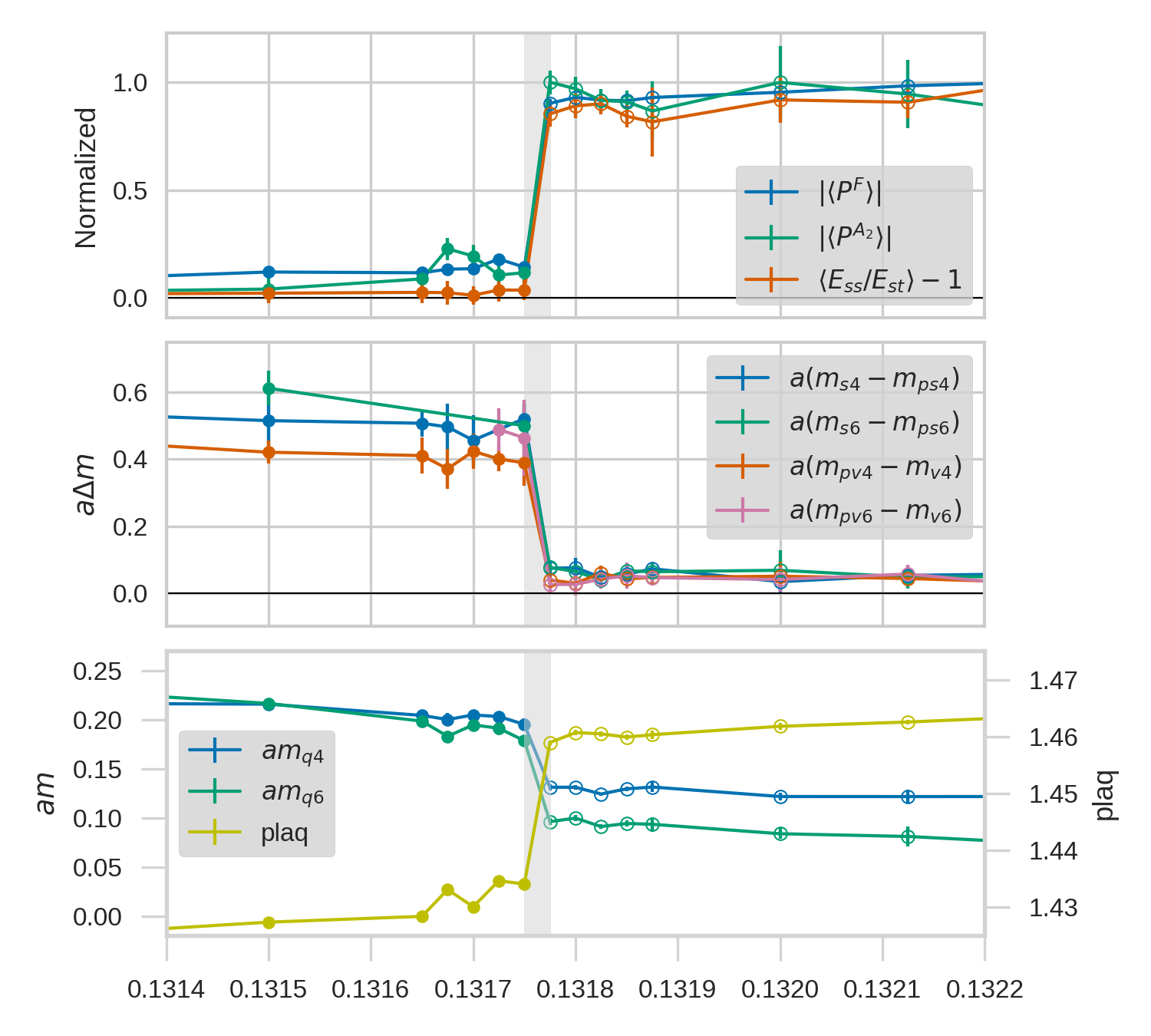}
	\caption{
		Slice varying $\kappa_6$ while holding $\beta=7.4$ and $\kappa_4=0.1285$ constant on $N_t=6$.
		The gray band brackets the transition.
		Closed (open) dots indicate that the ensemble is confined (deconfined) as shown by the long flow time Polyakov loops test \cite{other_thermo_proceedings}.
		\textbf{Top:}
			Polyakov loops for both irreps and the flow anisotropy observable \cite{Datta:2015bzm,Datta:2016kea,Wandelt:2016oym} respond simultaneously.
			Quantities are plotted as a proportion of their maximum value along the slice for ease of comparison of their qualitative behavior.
		\textbf{Center:}
			Mass splittings between parity partner mesonic states: scalar vs. pseudoscalar, and vector vs. pseudovector.
			The chiral transition occurs simultaneously for both irreps.
		\textbf{Bottom:}
			Lattice units quark masses for each irrep and plaquette.
			All quantities jump discontinuously at the transition.
	}
	\label{fig:mrep-k6-slice}
\end{figure}

The strong coupling phase of our system is confined and both representations of fermions
are chirally condensed.
Everywhere that we have investigated, we observe a single thermal phase transition at which both irreps deconfine and chiral symmetry for both species of fermion is restored.
Figure~\ref{fig:mrep-k6-slice} shows the behavior of all of our phase diagnostics on a typical
 slice through bare parameter space. It shows results versus $\kappa_6$ across the transition at
 fixed $\beta$ and $\kappa_4$.
Simultaneously, both Polyakov loops acquire large expectation values and chiral doubling sets 
in, indicating that the chiral transitions and confinement transitions for each irrep all coincide.

\begin{figure}
	\centering
	\includegraphics[width=5in,clip,trim={0in 0.75in 0in 0.75in}]{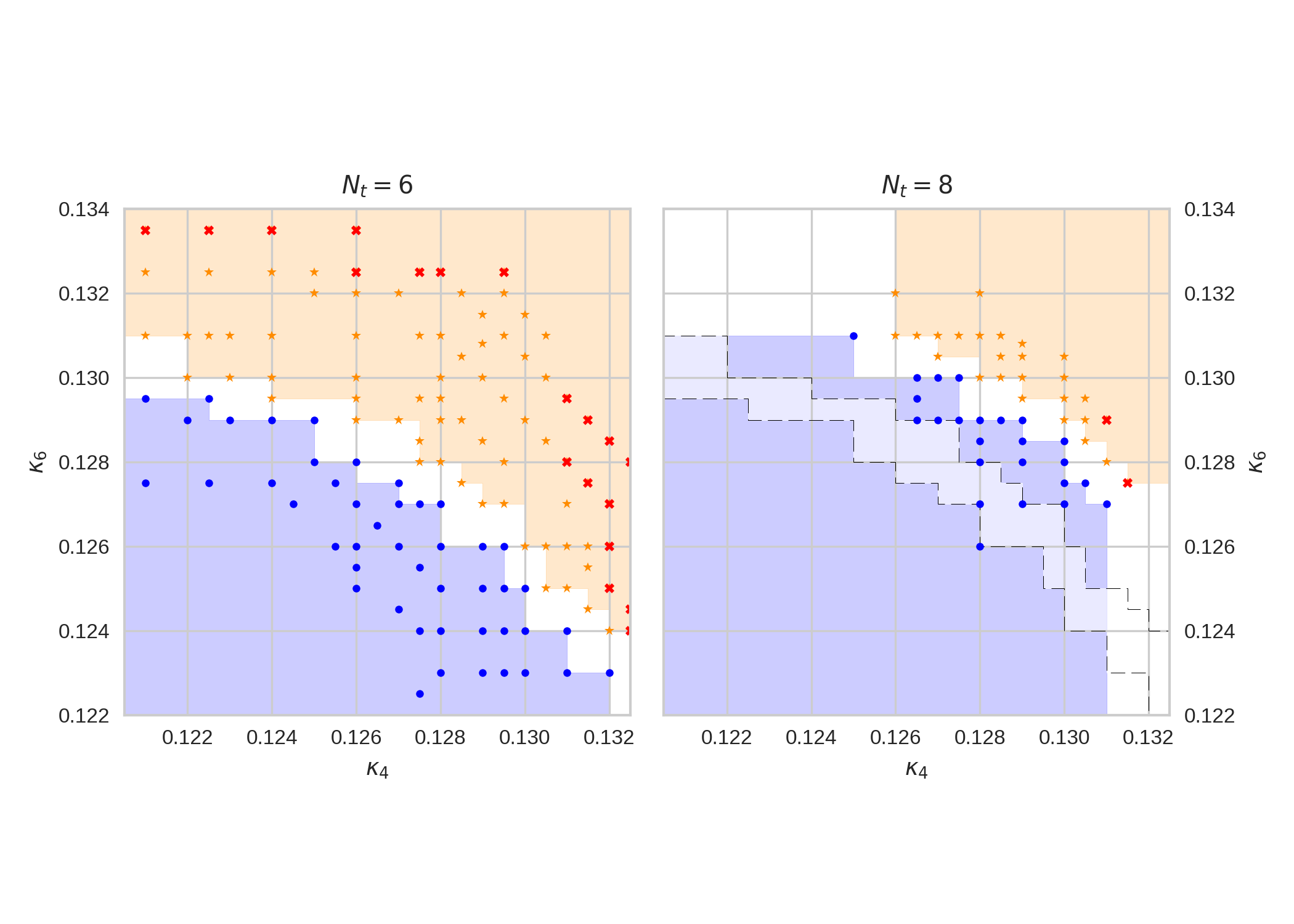}
	\caption{
		Phase diagram in the multirep theory at constant $\beta=7.75$.
		At left is the diagram for $N_t=6$ lattices, while at right is the same region of bare parameter space for $N_t=8$ lattices.
		Blue dots indicate confined ensembles with $m_q>0$ for both species.
		Yellow stars (red Xs) indicate deconfined ensembles with $m_q>0$ for both ($m_q<0$ for either) species.
		Blue regions are unambiguously confined, orange regions are unambiguously deconfined, and white regions are phase-ambiguous (transition lives here).
		The pale band overlaid on $N_t=8$ is the phase-ambiguous region from $N_t=6$, demonstrating that the transition moves as $N_t$ is varied.
	}
	\label{fig:mrep-map-775}
\end{figure}

In order to diagnose whether the transition encountered is a bulk transition, we have 
investigated two different temporal extents, $N_t=6$ and $N_t=8$.
Figure~\ref{fig:mrep-map-775} shows the resulting two $\kappa_4-\kappa_6$ phase diagrams for $\beta=7.75$.
The phase-ambiguous region from the $N_t=6$ diagram is overlaid on to the $N_t=8$ diagram, 
demonstrating that the transition moves substantially as $N_t$ is changed.
This strong response to the change in temperature indicates the transition is a physical thermal transition.
Decreasing the bare temperature $1/N_t$ at fixed bare parameters (and thus fixed $a$) decreases the physical temperature $T=1/a N_t$.
At lower physical temperature, the deconfinement transition will occur at lighter quark masses and thus greater $\kappa$s for a given $\beta$.

\begin{figure}
	\centering
	\sidecaption
	\includegraphics[width=3in,clip,trim=0.1in 0 0.45in 0]{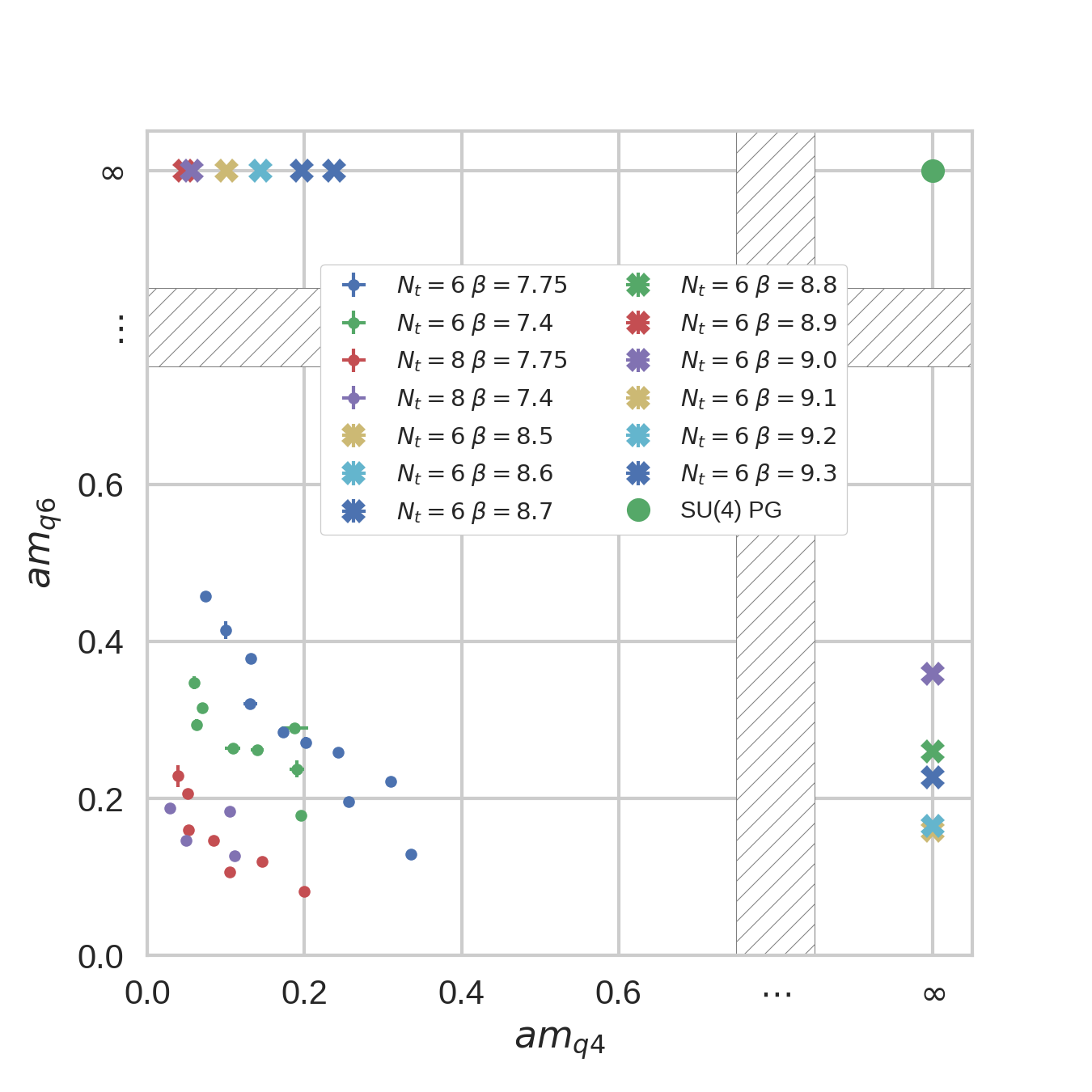}
	\caption{
		Multirep Columbia plot, by analogy with QCD \cite{Petreczky:2012rq}.
		Each point is an ensemble on the transition boundary, found by 
		varying $\kappa$s at some fixed $\beta$ and $N_t$.
		Each color is associated with a different $\beta$ and $N_t$; 
		Circles indicate the transition encountered near this ensemble is first-order.
		Xs indicate the transition found is continuous.
		All quark masses are plotted in lattice units.
		Pure gauge SU(4) is known to be first order \cite{Datta:2009jn}.
	}
	\label{fig:awi-columbia-plot}
\end{figure}

The observed transition appears to be strongly first order.
As can be seen in figure~\ref{fig:mrep-k6-slice}, all observables that we have investigated
 jump discontinuously at the phase transition.
Further, we have observed several tunneling events during equilibration, indicating
 strong metastability in HMC time.
These events can occur when we start equilibrating an ensemble in one phase using a
 configuration from an ensemble in the other phase as a seed.
After $O(10-100)$ times longer than typical equilibration times for these lattices
(sometimes after 1000 HMC trajectories), the system will suddenly tunnel to the target
 phase and all observables acquire values typical for the new phase.
This is characteristic behavior for a first-order transition.
The quark masses where we have investigated the transition are shown in figure~\ref{fig:awi-columbia-plot}.

\section{Continuum Theory: Multirep Pisarski-Wilczek}
\label{sec:pisarski-wilczek}
To see whether the first order behavior we observe is an expected result, we performed a Pisarski-Wilczek \cite{Pisarski:1983ms} stability analysis of the
chiral transition in the multirep theory.
The calculation is an analysis of the critical behavior of an effective three-dimensional field theory of the two chiral condensates of the theory.
If any infrared-stable fixed points exist for this theory, the transition can be second order if it occurs in the basin of attraction of that fixed point (but may still be first order otherwise).
If no infrared-stable fixed points exist, the transition must be first order.
To proceed, we first identify the symmetries and spontaneous symmetry-breaking (SSB) pattern of our theory 
and use them to construct the most general Landau-Ginzburg-Wilson (LGW) Lagrangian, including 
only relevant and marginal terms.
We compute the $\beta$ functions for this effective theory to one loop using the
 $\epsilon$ expansion: we expand as usual in small $\epsilon = 4-d$, but then set $\epsilon=1$.
Finally, we perform a stability analysis of the resulting $\beta$ functions in the zero-mass
 theory (with relative signs of couplings constrained to induce the correct SSB pattern).
The validity of the calculation is limited by both the $\epsilon$ expansion and by working to only one loop.
Analyses which go to higher order in perturbation theory and treat three-dimensionality more cautiously can find stable fixed points that are missed by the approach used here \cite{Basile:2005hw,Basile:2004wa}.

We begin by considering $N_\rlF$ Dirac flavors of
 \rF fermions and $N_\rlA$ Dirac flavors of \rA fermions.
Theories with fermions charged under $n$ different irreps generically have $n$ independent axial 
symmetries, which can be used to construct $n-1$ non-anomalous axial symmetries.
For our theory, there are two independent $\mathrm{U}(1)_A$s.
There is a non-anomalous linear combination of these two axial currents.
Thus, there is a good $\mathrm{U}(1)_A$ which is spontaneously broken, yielding a flavor-singlet axial pNGB.
This $\mathrm{U}(1)_A$ must be a good symmetry of our effective theory.
\cite{Clark:1986vk,DeGrand:2016pgq} 

The \rF of SU(4) is complex and so the \rF sector has the typical chiral
 symmetry ${\mathrm{SU}(N_\rlF)_L \times \mathrm{SU}(N_\rlF)_R}$.
The \rA of SU(4) is a real irrep, which expands the \rA-sector chiral symmetry
 group to $\mathrm{SU}(2 N_\rlA)$ \cite{Peskin:1980gc}.
The SSB pattern of our theory is then
\begin{equation}
\mathrm{SU}(N_\rlF)_L \times \mathrm{SU}(N_\rlF)_L
\times \mathrm{SU}(2 N_\rlA) \times \mathrm{U}(1)_A
\rightarrow
\mathrm{SU}(N_\rlF)_V \times \mathrm{SO}(2 N_\rlA).
\end{equation}

For the fundamental fermion fields $q$, define the complex $N_\rlF \times N_\rlF$ matrix
 field $\phi \sim \bar{q}^i_R q^{}_{Lj}$ (where $i$ and $j$ are Dirac flavor indices).
It transforms under chiral rotations like
\begin{equation}
\phi \rightarrow e^{2 i \alpha_\rlF} U_L \phi U_R,
\label{eqn:f-transf}
\end{equation}
where $U_L, U_R \in \textrm{SU}(N_\rlF)$ and $\alpha_\rlF$ is the angle of the axial rotation for the \rFs \cite{Pisarski:1983ms}.
For the two-index antisymmetric fermion fields $Q$, define the symmetric
 complex $2 N_\rlA \times 2 N_\rlA$ field $\theta \sim Q_I Q_J$ (where $Q$ is a left-handed Weyl field, and $I$ and $J$ are Weyl flavor indices). It  transforms under chiral rotations like
\begin{equation}
\theta \rightarrow e^{2 i \alpha_\rlA} V^T \theta V,
\label{eqn:a2-transf}
\end{equation}
where $V \in \textrm{SU}(2 N_\rlA)$ and $\alpha_\rlA$ is the angle of the axial rotation \cite{Basile:2004wa,Basile:2005hw}.

To construct the LGW Lagrangian, we write down all non-irrelevant terms that are
 invariant under equations (\ref{eqn:f-transf}) and (\ref{eqn:a2-transf}).
Including the single non-irrelevant term which couples \rF and \rA fermions, the full Lagrangian is
\begin{equation}
\begin{split}
\mathcal{L}^\text{Multirep}	
& = \Tr [\partial_\mu \phi^\dagger \partial^\mu \phi ]
+ r_\rlF \Tr [ \phi^\dagger \phi ]
+ u_\rlF ( \Tr [ \phi^\dagger \phi ] )^2
+ v_\rlF \Tr [ (\phi^\dagger \phi)^2 ]
\\
& + \Tr [\partial_\mu \theta^\dagger \partial^\mu \theta ]
+ r_\rlA \Tr [ \theta^\dagger \theta ]
+ u_\rlA ( \Tr [ \theta^\dagger \theta ] )^2
+ v_\rlA \Tr [ (\theta^\dagger \theta)^2 ]
\\
& + w \Tr [ \phi^\dagger \phi ] \Tr [ \theta^\dagger \theta ]
.
\label{eqn:L-multirep}
\end{split}
\end{equation}

In order to break the axial symmetry of the theory from
 $\mathrm{U}(1)_A^{(\rlF)} \times \mathrm{U}(1)_A^{(\rlA)}$ to the non-anomalous
 $\mathrm{U}(1)_A$, we must introduce all non-irrelevant terms constructed from determinants,
which respect the unbroken $\mathrm{U}(1)_A$.
They are
${\det{\phi} \rightarrow e^{2 i N_\rlF \alpha_\rlF} \det{\phi}}$
and
${\det{\theta} \rightarrow e^{4 i N_\rlA \alpha_\rlA} \det{\theta}}$.
For the exact flavor content of our theory ($N_\rlF=N_\rlA=2$), we find
 that simultaneous axial rotations obeying $\alpha_\rlF = - 2 \alpha_\rlA$ are unbroken 
symmetries \cite{DeGrand:2016pgq}.
The lowest-order such term that respects this symmetry is
$\delta \mathcal{L} \sim \det \phi \det \theta + \text{(c.c.)}$,
but because $[\det \phi] = N_\rlF = 2$ and $[\det \theta] = 2 N_\rlA = 4$, this term
 is irrelevant for our theory.
Thus, equation~\ref{eqn:L-multirep} is the final Lagrangian, with both $\textrm{U}(1)_A$s 
as good symmetries.
Physically, we find that the axial anomaly does not play a role in the critical behavior of
 the theory.

We observe only one transition in our data,
 so we only consider the behavior of the theory when both irreps are simultaneously driven to
 criticality; thus, we set $r_\rlF=r_\rlA=0$.
We then compute the $\beta$ functions of the resulting Lagrangian to one loop.
With all of the couplings redefined by the same overall factor to absorb geometric constants, these are
\begin{alignat*}{4}
&\beta_{u_\rlF} &=& -u_\rlF &&
+ (N_\rlF^2 + 4) u_\rlF^2
+ 4 N_\rlF u_\rlF v_\rlF
+ 3 v_\rlF^2 
+ 4 N_\rlA (2 N_\rlA + 1) w^2
\\
&\beta_{v_\rlF} &=& -v_\rlF &&
+ 6 u_\rlF v_\rlF
+ 2 N_\rlF v_\rlF^2
\\
&\beta_{u_\rlA} &=& -u_\rlA &&
+ \tfrac{1}{2}(4 N_\rlA^2 + 2 N_\rlA + 8) u_\rlA^2
+ 2 (2 N_\rlA + 1) u_\rlA v_\rlA
+ \tfrac{3}{2} v_\rlA^2
+ 4 N_\rlF^2 w^2
\\
&\beta_{v_\rlA} &=& -v_\rlA &&
+ 6 u_\rlA v_\rlA
+ (2 N_\rlA + \tfrac{5}{2}) v_\rlA^2
\\
&\beta_{w}      &=& -w &&
+ w \left(
(N_\rlF^2 + 1) u_\rlF
+ 2 N_\rlF v_\rlF
+ \tfrac{1}{2} (4 N_\rlA^2 + 2 N_\rlA + 4) u_\rlA
+ (2 N_\rlA + 1) v_\rlA
+ 2 w
\right).
\label{eqn:pw-beta-fns}
\end{alignat*}
A stability analysis amounts to finding the fixed points of these $\beta$ 
functions and determining whether any of the eigenvalues of the stability matrix 
 $\partial \beta_{g_i} / \partial g_j$  (where $g_i \in \{u_\rlF, v_\rlF, u_\rlA, v_\rlA, w\}$) are
 negative there.
We find six fixed points, none of which are infrared-stable.
Our calculation thus indicates that the transition in our theory should be first-order.
This result is consistent with our data.

\section{Conclusion}
\label{sec:conclusion}

We do not observe any separation of phases: our lattice data indicate that both irreps
 confine and break chiral symmetry simultaneously. Our system does not show scale separation.
Further, this transition appears to be strongly first-order, consistent with
 our multirep Pisarski-Wilczek calculation.

Although the finite-temperature properties of this model are not  relevant for
 LHC phenomenology, they may have  implications for cosmology.
In particular, a first-order phase transition in the early universe would be expected to 
generate a primordial gravitational wave signature, see e.g.~\cite{Schwaller:2015tja, Caprini:2015zlo}.

\subsection*{Acknowledgements}

\footnotesize
Research was supported by U.S.~Department of Energy Grant
Number  under grant DE-SC0010005 (Colorado)
and  by the Israel Science Foundation
under grant no.~449/13 (Tel Aviv).  Brookhaven National Laboratory is supported
by the U.~S.~Department of Energy under contract DE-SC0012704.
This work utilized the Janus supercomputer, which is supported by the National Science Foundation (award number CNS-0821794) and the University of Colorado Boulder. The Janus supercomputer is a joint effort of the University of Colorado Boulder, the University of Colorado Denver and the National Center for Atmospheric Research.
Additional computations were done on facilities of the USQCD Collaboration at Fermilab,
which are funded by the Office of Science of the U.~S. Department of Energy.
The computer code is based on the publicly available package of the
 MILC collaboration~\cite{MILC}. 

\clearpage

\bibliography{lattice2017}

\begin{thebibliography}{30}

\bibitem{Raby:1979my}
S.~Raby, S.~Dimopoulos, L.~Susskind, Nucl. Phys. \textbf{B169}, 373 (1980)

\bibitem{Kogut:1983sm}
J.B. Kogut, M.~Stone, H.W. Wyld, S.H. Shenker, J.~Shigemitsu, D.K. Sinclair,
  Nucl. Phys. \textbf{B225}, 326 (1983)

\bibitem{Kogut:1982rt}
J.B. Kogut, M.~Stone, H.W. Wyld, W.R. Gibbs, J.~Shigemitsu, S.H. Shenker, D.K.
  Sinclair, Phys. Rev. Lett. \textbf{50}, 393 (1983)

\bibitem{Kogut:1984nq}
J.B. Kogut, J.~Shigemitsu, D.K. Sinclair, Phys. Lett. \textbf{138B}, 283 (1984)

\bibitem{Kogut:1984sb}
J.B. Kogut, J.~Shigemitsu, D.K. Sinclair, Phys. Lett. \textbf{145B}, 239 (1984)

\bibitem{Karsch:1998qj}
F.~Karsch, M.~Lutgemeier, Nucl. Phys. \textbf{B550}, 449 (1999),
  \texttt{hep-lat/9812023}

\bibitem{DeGrand:2015zxa}
T.~DeGrand, Rev. Mod. Phys. \textbf{88}, 015001 (2016), \texttt{1510.05018}

\bibitem{ben_review}
B.~Svetitsky, \emph{{Looking behind the Standard Model with lattice gauge
  theory}}, in \emph{Proceedings,
  \href{http://inspirehep.net/record/1425631}{35th International Symposium on
  Lattice Field Theory (Lattice2017)}: Granada, Spain}, to appear in EPJ Web
  Conf., \texttt{1708.04840}

\bibitem{spectro_paper}
V.~Ayyar, T.~DeGrand, M.~Golterman, D.C. Hackett, W.I. Jay, E.T. Neil,
  Y.~Shamir, B.~Svetitsky, work in progress

\bibitem{DeGrand:2014rwa}
T.~DeGrand, Y.~Shamir, B.~Svetitsky, Phys. Rev. \textbf{D90}, 054501 (2014),
  \texttt{1407.4201}

\bibitem{Hasenfratz:2001hp}
A.~Hasenfratz, F.~Knechtli, Phys. Rev. \textbf{D64}, 034504 (2001),
  \texttt{hep-lat/0103029}

\bibitem{Hasenfratz:2007rf}
A.~Hasenfratz, R.~Hoffmann, S.~Schaefer, JHEP \textbf{05}, 029 (2007),
  \texttt{hep-lat/0702028}

\bibitem{Bernard:1999kc}
C.W. Bernard, T.A. DeGrand, Nucl. Phys. Proc. Suppl. \textbf{83}, 845 (2000),
  \texttt{hep-lat/9909083}

\bibitem{Shamir:2010cq}
Y.~Shamir, B.~Svetitsky, E.~Yurkovsky, Phys. Rev. \textbf{D83}, 097502 (2011),
  \texttt{1012.2819}

\bibitem{other_thermo_proceedings}
V.~Ayyar, D.C. Hackett, W.I. Jay, E.T. Neil, Y.~Shamir, B.~Svetitsky (TACO),
  \emph{{Confinement study of an SU(4) gauge theory with fermions in multiple
  representations}}, in \emph{Proceedings,
  \href{http://inspirehep.net/record/1425631}{35th International Symposium on
  Lattice Field Theory (Lattice2017)}: Granada, Spain}, to appear in EPJ Web
  Conf.

\bibitem{DeGrand:2015lna}
T.~DeGrand, Y.~Liu, E.T. Neil, Y.~Shamir, B.~Svetitsky, Phys. Rev.
  \textbf{D91}, 114502 (2015), \texttt{1501.05665}

\bibitem{Datta:2015bzm}
S.~Datta, S.~Gupta, A.~Lytle, Phys. Rev. \textbf{D94}, 094502 (2016),
  \texttt{1512.04892}

\bibitem{Datta:2016kea}
S.~Datta, S.~Gupta, A.~Lytle, PoS \textbf{LATTICE2016}, 091 (2016),
  \texttt{1612.07985}

\bibitem{Wandelt:2016oym}
M.~Wandelt, F.~Knechtli, M.~Günther, JHEP \textbf{10}, 061 (2016),
  \texttt{1603.05532}

\bibitem{Basile:2004wa}
F.~Basile, A.~Pelissetto, E.~Vicari, JHEP \textbf{02}, 044 (2005),
  \texttt{hep-th/0412026}

\bibitem{Basile:2005hw}
F.~Basile, A.~Pelissetto, E.~Vicari, PoS \textbf{LAT2005}, 199 (2006),
  \texttt{hep-lat/0509018}

\bibitem{Petreczky:2012rq}
P.~Petreczky, J. Phys. \textbf{G39}, 093002 (2012), \texttt{1203.5320}

\bibitem{Datta:2009jn}
S.~Datta, S.~Gupta, Phys. Rev. \textbf{D80}, 114504 (2009), \texttt{0909.5591}

\bibitem{Pisarski:1983ms}
R.D. Pisarski, F.~Wilczek, Phys. Rev. \textbf{D29}, 338 (1984)

\bibitem{Clark:1986vk}
T.E. Clark, C.N. Leung, S.T. Love, J.L. Rosner, Phys. Lett. \textbf{B177}, 413
  (1986)

\bibitem{DeGrand:2016pgq}
T.~DeGrand, M.~Golterman, E.T. Neil, Y.~Shamir, Phys. Rev. \textbf{D94}, 025020
  (2016), \texttt{1605.07738}

\bibitem{Peskin:1980gc}
M.E. Peskin, Nucl. Phys. \textbf{B175}, 197 (1980)

\bibitem{Schwaller:2015tja}
P.~Schwaller, Phys. Rev. Lett. \textbf{115}, 181101 (2015), \texttt{1504.07263}

\bibitem{Caprini:2015zlo}
C.~Caprini et~al., JCAP \textbf{1604}, 001 (2016), \texttt{1512.06239}

\bibitem{MILC}
{MILC Collaboration},
  \urlstyle{tt}\url{http://www.physics.utah.edu/~detar/milc/}

\end{thebibliography}

\end{document}